\documentclass[final,5p,times,twocolumn]{elsarticle}

\usepackage{lineno}
\usepackage[hidelinks]{hyperref}

\usepackage{graphicx}
\usepackage{amsmath}   
\def\pmbanner{{\hrule height 1 pt}\vskip35pt{NIMA POST-PROCESS BANNER TO BE REMOVED AFTER FINAL ACCEPTANCE}\vskip35pt{\hrule height 4pt}\vskip20pt}

\begin{document}

\begin{frontmatter}

\title{\pmbanner Study of environment friendly gas mixtures for the Resistive Plate Chambers of the ATLAS phase-2 upgrade}

\author[1]{G.Proto\corref{cor1}%
\fnref{fn1}  on behalf of the ATLAS Muon Community}

\ead{gproto@cern.ch}




\affiliation[1]{organization={Max Planck Institute for Physics}, 
                 addressline={Boltzmannstr. 8},
                 postcode={85748}, 
                 city={Munich}, 
                 country={Germany}}

\begin{abstract}
The standard gas mixture for the Resistive Plate Chambers (RPC), composed of $\rm{C_{2}H_{2}F_{4}/i-C_{4}H_{10}/SF_{6}}$, allows the detector operation in avalanche mode, as required by the high-luminosity collider experiments. The gas density, the low current and the comfortable avalanche-streamer separation guarantee high detection efficiency, rate capability and slow detector ageing. However, the mixture has a high Global Warming Potential (GWP $\rm{\sim}$1430), primarily due to the presence of $\rm{C_{2}H_{2}F_{4}}$. The $\rm{C_{2}H_{2}F_{4}}$ and $\rm{SF_{6}}$ are not recommended for industrial uses anymore, thus their availability will be increasingly difficult over time and the search for an alternative gas mixture is then of absolute priority within the RPC community. CERN is also driving efforts to reduce these gases, as they contribute significantly to the LHC greenhouse gas emissions. Within the ATLAS experiment, the search for an environment-friendly gas mixture involves both the legacy system and the new generation of RPC detectors for the HL-LHC [1]. The thin 1 mm gas gap of the latter requires a high-density in order to achieve high efficiency, due to the less active target available for the primary ionization. The mixture should also guarantee good timing performance and ensure the detector longevity compared with the standard one. In this paper, the results obtained on a RPC operated with alternative gas mixtures are shown, following two different approaches. The first study consists of the replacement of the $\rm{C_{2}H_{2}F_{4}}$ with a mixture of $\rm{C_{3}H_{2}F_{4}/CO_{2}}$, which has a significantly lower Global Warming Potential (GWP$\rm{\sim}$ 200). In order to achieve an efficiency greater than 90\%, the concentration of $\rm{C_{2}H_{2}F_{4}}$ must be maintained above 50\%. Meanwhile, the addition of $\rm{CO_{2}}$, improves the time resolution, reaching approximately 285 ps. The second approach consists in adding a modest fraction of $\rm{CO_{2}}$ in the standard gas, with the aim to reduce the $\rm{C_{2}H_{2}F_{4}}$ emissions. In this case, the currents are lower compared to those observed with the $\rm{C_{3}H_{2}F_{4}/CO_{2}}$-based gas mixtures, and no significant impact on detector aging is expected, as the overall composition of the mixture remains similar to the standard gas. In both approaches, the concentration of $\rm{SF_6}$  must be maintained at a minimum of 1\% to prevent premature streamer formation and the associated increase in current. The paper provides a detailed study of efficiency, time resolution, and current under different irradiation backgrounds. 
\end{abstract}


\end{frontmatter}

\section{Introduction}
This study explores two approaches to reduce the environmental impact of the RPC gas mixture. The first approach involves replacing $\rm{C_{2}H_{2}F_{4}}$ with a $\rm{C_{3}H_{2}F_{4}/CO_{2}}$-based mixture, which has a significantly lower GWP ( $\sim$ 200, to be compared to 1430 of the standard gas). The second approach introduces a modest fraction of $\rm{CO_{2}}$ (in the range 30\%-40\%) into the standard gas mixture, aiming to reduce $\rm{C_{2}H_{2}F_{4}}$ emissions without negatively impacting the detector aging. Tests are performed at the Gamma Irradiation Facility (GIF++) at CERN on a production ATLAS-BIS78-phase-1 upgrade detector [1] and a small prototype with the same layout. The study presents a comprehensive evaluation of active target (efficiency), time resolution, and current in different irradiation environments. The gas mixtures have been studied under both LHC and HL-LHC $\gamma$-background, corresponding to 100 Hz/$\rm{cm^{2}}$ and 200 Hz/$\rm{cm^{2}}$ respectively, as well as at higher radiation levels. 
\section{Operating the RPC with $\rm{C_{3}H_{2}F_{4}/CO_{2}}/$i-$\rm{C_{4}H_{10}/SF_{6}}$ gas mixtures}
\label{sec1}
\begin{figure}[t]
\centering
\includegraphics[width=0.53\textwidth]{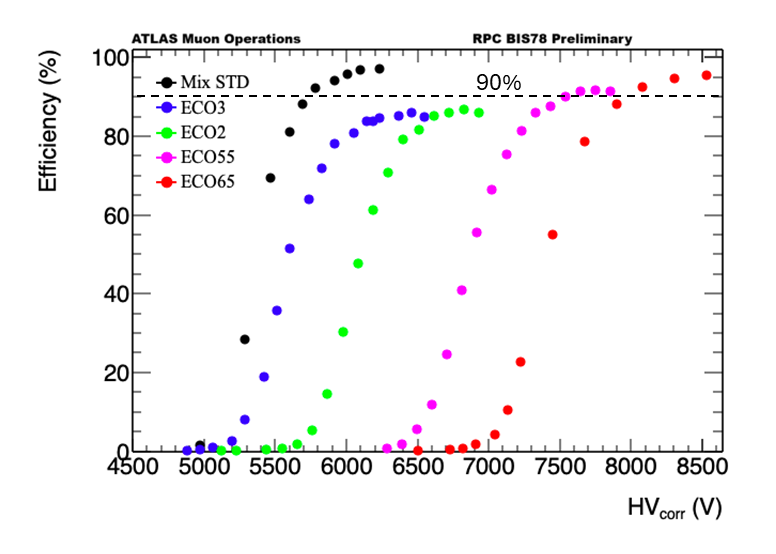}
\caption{Efficiency as a function of the high voltage comparing the standard gas mixture (black) with ECO3 (blue), ECO2 (green), ECO55 (pink) and ECO65 (red) [2].
}\label{fig1}
\end{figure}

\begin{figure}[t]
\centering
\includegraphics[width=0.5\textwidth]{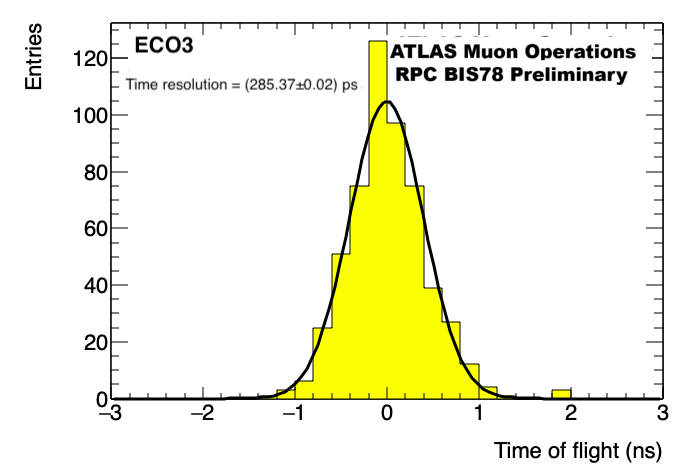}
\caption{Time resolution of the 1 mm RPC operated with ECO3 gas mixture [2].
}\label{fig2}
\end{figure}

\begin{figure}[t]
\centering
\includegraphics[width=0.5\textwidth]{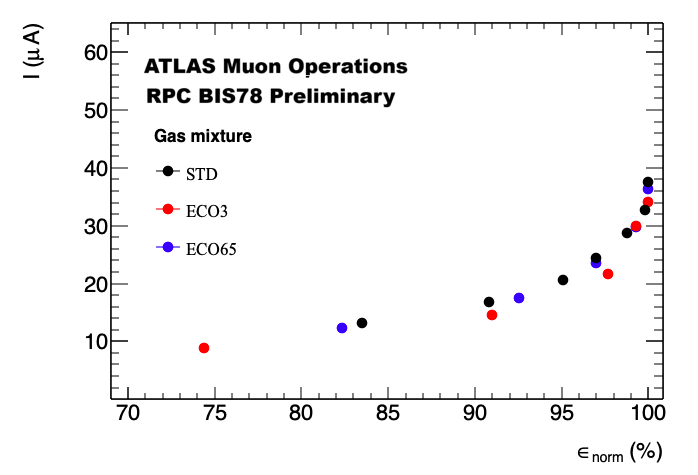}
\caption{Photon current as a function of the normalized efficiency comparing the standard gas mixture (black) with ECO3 (red) and ECO65 (blue). The efficiency is normalized to the maximum plateau efficiency of the gas mixture itself.
}\label{fig3}
\end{figure}

In this section the results obtained operating the RPC with gas mixtures composed of $\rm{C_{3}H_{2}F_{4}/CO_{2}}/$i-$\rm{C_{4}H_{10}/SF_{6}}$ are shown [2].  The gas mixtures studied are :
\begin{itemize}
  \item ECO3=$\rm{25\%C_{3}H_{2}F_{4}/70\%CO_{2}}/$4\%i-$\rm{C_{4}H_{10}/1\%SF_{6}}$;
  \item ECO2=$\rm{35\%C_{3}H_{2}F_{4}/60\%CO_{2}}/$4\%i-$\rm{C_{4}H_{10}/1\%SF_{6}}$;
  \item ECO55=$\rm{55\%C_{3}H_{2}F_{4}/40\%CO_{2}}/$4\%i-$\rm{C_{4}H_{10}/1\%SF_{6}}$;
  \item ECO65=$\rm{65\%C_{3}H_{2}F_{4}/30\%CO_{2}}/$4\%i-$\rm{C_{4}H_{10}/1\%SF_{6}}$.
\end{itemize}
The results have been compared with those obtained with the standard gas mixture, composed of 95\%$\rm{C_{2}H_{2}F_{4}/4.7\%i-C_{4}H_{10}/0.3\%SF_{6}}$.\\
The efficiency as a function of the high voltage is shown in Figure \ref{fig1}. The plateau efficiency achieved with ECO2 and ECO3 gas mixtures is below 90\%, while for ECO55 and ECO65 it exceeds 90\%. The reduced density of the $\rm{CO_{2}}$ results in a small active target available for the primary ionization. Consequently, the mixtures with higher $\rm{CO_{2}}$ concentrations, ECO2 and ECO3, exhibit low detection efficiency. On the other hand, ECO55 and ECO65 show high-efficiency plateau, above 95\%, due to the predominant presence of a high-density gas, the $\rm{C_{3}H_{2}F_{4}}$. The advantage to work with $\rm{CO_{2}}$-based gas mixtures lies in the improvement of the RPC time resolution, as shown in Figure \ref{fig2}. This is due to the higher drift speed in $\rm{CO_{2}}$ compared to $\rm{C_{2}H_{2}F_{4}}$. The time resolution achieved operating the RPC with ECO3 gas mixture is $\rm{285}$ ps, compared to $\rm{330}$ ps obtained with the standard gas [2].\\ One of the most critical point concerning the usage of the $\rm{C_{3}H_{2}F_{4}}$-based gas mixtures is the detector aging effect, that is currently under study [3]. The double Carbon-Carbon bond characteristic of $\rm{C_{3}H_{2}F_{4}}$ makes the breakage of the molecule easier, potentially resulting in a larger production of fluorine radicals compared to the standard gas, that can damage the electrodes surfaces in long term operation. This effect can be mitigated by using a low FE electronics threshold, as shown in Figure \ref{fig3}. It shows the photon current (main ATLAS cavern background in RPC) as a function of the normalized efficiency, comparing the standard gas with ECO3 and ECO65 with the RPC operated at 1 fC threshold. The plot shows that the photon contribution to the current is independent from the gas mixture  at the same normalized efficiency. This result is very promising because it suggests that working at 1 fC threshold, these eco-friendly gases might ensure the same aging and the same rate capability of the standard gas. Moreover, the lower is the threshold the lower the electric field at maximum efficiency. Consequently, the probability of producing Fluorine radicals is also reduced, allowing the usage of gas mixtures with high $\rm{C_{3}H_{2}F_{4}}$ content, like ECO65, not possible otherwise.

\begin{figure}[!h!t]
\centering
\includegraphics[width=0.53\textwidth]{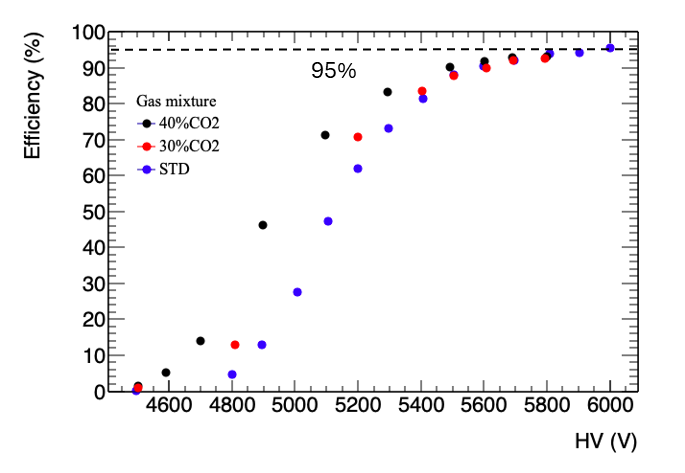}
\caption{Efficiency as a function of the high voltage comparing the standard gas (blue) with the gas mixtures containing 40\% $\rm{CO_{2}}$ (black) and 30\% $\rm{CO_{2}}$ (red) [4]}\label{fig4}
\end{figure}
\begin{figure}[!h!t]
\centering
\includegraphics[width=0.5\textwidth]{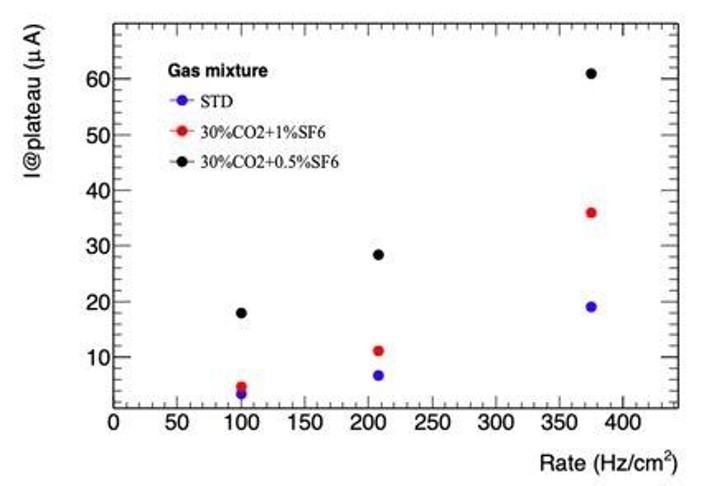}
\caption{RPC current measured at 94\% efficiency as a function of the photon rate, comparing the standard gas (blue), the mixture containing (30\% CO2 + 1\% SF6) (red) and the one composed of (30\% CO2 + 0.5\% SF6) (black) [4]}\label{fig5}
\end{figure}
\section{Operating the RPC with $\rm{C_{2}H_{2}F_{4}/CO_{2}}/$i-$\rm{C_{4}H_{10}/SF_{6}}$ gas mixtures}
\label{sec2}
The second possibility to reduce the GWP of the standard gas and reduce the $\rm{C_{2}H_{2}F_{4}}$ emissions, is to add a certain amount of  $\rm{CO_{2}}$ in the standard gas itself. The following gas mixtures [4], with a GWP $\rm{\sim}$1100, have been studied:
\begin{itemize}
  \item $\rm{65\%C_{2}H_{2}F_{4}/30\%CO_{2}}/$4\%i-$\rm{C_{4}H_{10}/1\%SF_{6}}$;
  \item $\rm{55\%C_{2}H_{2}F_{4}/40\%CO_{2}}/$4\%i-$\rm{C_{4}H_{10}/1\%SF_{6}}$;
  \item $\rm{65.5\%C_{2}H_{2}F_{4}/30\%CO_{2}}/$4\%i-$\rm{C_{4}H_{10}/0.5\%SF_{6}}$.
\end{itemize}
Figure \ref{fig4} shows the efficiency as a function of the high voltage comparing the standard gas with gas mixtures containing 40\%$\rm{CO_{2}}$ and 30\%$\rm{CO_{2}}$. The plot shows that there is no significantly degradation in terms of efficiency, being 96\% for the standard gas and 94\% for the alternative gas mixtures. The trend of the current under irradiation shows that the mixture containing 40\%$\rm{CO_{2}}$ has a current significantly higher with respect to the standard gas (1.7 times higher), while the mixture with 30\%$\rm{CO_{2}}$ shows a current 1.5 times higher [4]. This is due to the premature appearance of streamers occurring in the former case. In conclusion, the gas mixture with the addition of 30\%$\rm{CO_{2}}$ exhibits the best performance. In order to decrease the GWP of the gas mixture, the possibility to reduce the $\rm{SF_{6}}$ from 1\% to 0.5\% has been studied. The results are shown in Figure \ref{fig5} that shows the current at plateau as a function of the photon rate. Focusing on the rate of 200 Hz/$\rm{cm^{2}}$ that is the one expected during the High Luminosity phase of LHC, halving the $\rm{SF_{6}}$ concentration results in a current under irradiation 3 times higher, due to the premature appearance of streamers. This suggests that is not possible to reduced the $\rm{SF_{6}}$ concentration in these gas mixtures.

\section{Conclusions}
\label{sec3}
In this work, the performance of the new generation of ATLAS RPCs, designed for the High-Luminosity phase of the LHC (with a 1 mm gas gap width), operating with alternative gas mixtures, has been studied. The gas mixtures studied are composed of $\rm{C_{3}H_{2}F_{4}/CO_{2}}/$i-$\rm{C_{4}H_{10}/SF_{6}}$ (GWP$\rm{\sim}$ 200) and $\rm{C_{2}H_{2}F_{4}/CO_{2}}/$i-$\rm{C_{4}H_{10}/SF_{6}}$(GWP $\rm{\sim}$ 200). In the former case, the RPC shows very good performance in terms of efficiency (above 90\%) for $\rm{C_{3}H_{2}F_{4}}$ concentration above 50\% and an excellent time resolution of 285 ps (24\% better with respect to the standard gas [5]). The aging tests are ongoing in order to certificate the gas mixture for long term operation in the HL-LHC photon background environment.\\The gas mixture composed of $\rm{65\%C_{2}H_{2}F_{4}/30\%CO_{2}}/$4\%i-$\rm{C_{4}H_{10}/1\%SF_{6}}$ shows good performance in terms of efficiency and current under irradiation and is currently used in the RPC system of the ATLAS experiment since 1 year.\\ The possibility to further reduce the GWP will be studied in the near future, by replacing the $\rm{SF_{6}}$  with alternative low-GWP gases.


  \bibliographystyle{elsarticle-num-names} 
  
  \bibliography{bibfileTemplate}
[1] ATLAS Collaboration, Technical Design Report for the Phase-II Upgrade of the ATLAS Muon Spectrometer, CERN-LHCC-2017-017 \newline
[2] G.Proto, Study of environment-friendly gas mixtures for Resistive Plate Chambers in view of future applications, CERN-THESIS-2022-383\newline
[3] ECOGAS@Gif++ Collaboration, Preliminary results on long term operation of RPCs with eco-friendly gas mixtures under irradiation at the CERN Gamma Irradiation Facility, arXiv:2311.17574\newline
[4] G.Proto, Performance of new generation of Resistive Plate,Chambers operating with alternative gas mixtures, 3rd International Conference on Detector Stability and Aging Phenomena, Nucl.Instrum.Meth.A 1066 (2024),169580\newline
[5] G. Proto et al 2022 JINST 17 P05005

\end{document}